\documentclass[aps,prb,twocolumn,groupedaddress]{revtex4-2}
\usepackage{graphicx}
\usepackage{amssymb}
\usepackage{physics}
\usepackage{amsfonts}
\usepackage{hyperref}
\usepackage{cleveref}
\usepackage[caption=false]{subfig}
\usepackage[all]{hypcap}
\usepackage{xcolor}
\usepackage[normalem]{ulem}
\usepackage{comment}

\newcommand{\Ac}{\mathcal A}

\newcommand{\Hc}{\mathcal H}

\newcommand{\Oc}{\mathcal O}

\renewcommand{\l}{\lambda}
\newcommand{\s}{\sigma}
\renewcommand{\o}{\omega}

\newcommand{\Dc}{\mathcal{D}}

\begin{document}

\title{Integrability as an attractor of adiabatic flows}

\author{Hyeongjin Kim}
\email[]{hkim12@bu.edu}
\author{Anatoli Polkovnikov}
\affiliation{Department of Physics, Boston University, Boston, Massachusetts 02215, USA}

\date{\today}

\begin{abstract}
The interplay between quantum chaos and integrability has been extensively studied in the past decades. We approach this topic from the point of view of geometry encoded in the quantum geometric tensor, which describes the complexity of adiabatic transformations. In particular, we consider two generic models of spin chains that are parameterized by two independent couplings. In one, the integrability breaking perturbation is global while, in the other, integrability is broken only at the boundary. In both cases, the shortest paths in the coupling space lead towards integrable regions and we argue that this behavior is generic. These regions thus act as attractors of adiabatic flows similar to river basins in nature. Physically, the directions towards integrable regions are characterized by faster relaxation dynamics than those parallel to integrability, and the anisotropy between them diverges in the thermodynamic limit as the system approaches the integrable point.  We also provide evidence that the transition from integrable to chaotic behavior is universal for both models, similar to continuous phase transitions, and that the model with local integrability breaking quickly becomes chaotic but avoids ergodicity.
\end{abstract}

\maketitle

\section{Introduction}
There has been significant progress in understanding the nature of quantum chaos and integrability in the past few decades (see Refs.~\cite{Michael_Berry_1989,Haake1991,stockmann_1999,borgonovi2016quantum, d2016quantum} for review). On one hand, it is generally recognized that the random matrix behavior of quantum eigenstates [described by random matrix theory (RMT)] and energy spectrum are sensible measures of quantum chaos or, more accurately, quantum ergodicity, thermalization or mixing~\cite{RevModPhys.53.385,guhr1998random}. Emergent random matrix ensembles are connected with statistical mechanics and thermodynamics via the so-called eigenstate thermalization hypothesis (ETH)~\cite{Srednicki_1999,Deutsch_2018,d2016quantum}. On the other hand, Poisson statistics of the level spacings is considered a signature of quantum integrability through the Berry-Tabor conjecture~\cite{1977RSPSA.356..375B}. From the point of view of physical observables, integrable systems are generally nonergodic with their steady states being constrained by multiple conservation laws. In systems with local interactions, these states can be described by the so-called generalized Gibbs ensemble (GGE)~\cite{rigol2007relaxation}. In weakly nonintegrable systems, it is generally expected that these states first relax to such GGE states and then gradually relax to a true equilibrium. This slow relaxation mechanism is termed prethermalization~\cite{Berges_2004,Moeckel_2008,yurovsky2011memory, d2016quantum,durnin2021nonequilibrium}. In the context of weakly driven systems, some GGEs can be stabilized by weak driving, leading to, for example, robust turbulent cascades~\cite{gurarie1995field,Lenar_i__2018}.

A different approach for analyzing the transition between quantum chaos and integrability, which is more relevant to the present work, is based on the scaling analysis of the fidelity susceptibility $\chi$ as a function of integrability breaking perturbations~\cite{leblond2021universality,bulchandani2022onset,pandey2020adiabatic, surace2023weak, orlov2023adiabatic}. This approach was previously developed to study quantum phase transitions~\cite{zanardi2006ground,venuti2007quantum,kolodrubetz2013classifying}, allowing one to classify universal properties of quantum phases and phase transitions in an observable-independent way. When applied to excited states, fidelity susceptibility, together with other probes, enables one to identify the existence of a chaotic but nonergodic buffer region that generically separates integrable and ergodic/ETH regimes. This buffer zone is characterized by a stronger divergence of fidelity susceptibility with system size than in the ETH regime~\cite{pandey2020adiabatic,sels2021dynamical,leblond2021universality,bulchandani2022onset}. In a recent work, it was shown that a similar story holds for the emergence of chaos and ergodicity in classical Hamiltonian models with the fidelity susceptibility encoding the complexity of trajectory-preserving canonical transformations~\cite{lim2024defining}. Physically, this intermediate chaotic but nonergodic regime--both in quantum and classical systems--can be characterized by the emergence of diverging low-frequency tails of the spectral functions of observables, which are manifested by the slow relaxations of these observables to non-equilibrium states. It was also recognized that, at or near integrable points, there is a strong qualitative difference between the dynamics of integrability preserving and integrability breaking perturbations~\cite{pandey2020adiabatic,Zhang_2022}. This strong anisotropy motivates the consideration of the full structure of the quantum geometric tensor (QGT) in a parameter manifold containing more than one coupling. The QGT defines a natural Riemannian metric structure on the eigenstate manifolds ~\cite{provost1980riemannian,venuti2007quantum, kolodrubetz2013classifying}. While the definition of QGT sounds abstract, it naturally extends the notion of fidelity susceptibilities (which are diagonal components of the QGT). QGT is closely related to a range of physical phenomena like the long-time response of physical observables~\cite{kolodrubetz2017geometry}, quantum Fisher information~\cite{Helstrom1976quantum}, quantum speed limits~\cite{Funo_2017, Bukov_2019}, effective mass~\cite{kolodrubetz2017geometry}, superfluidity and superconductivity~\cite{peotta2023quantum}, and more (see Ref.~\cite{kolodrubetz2017geometry} for details).  

\begin{figure}[!htbp]
\centering
\includegraphics[width=8.6cm]{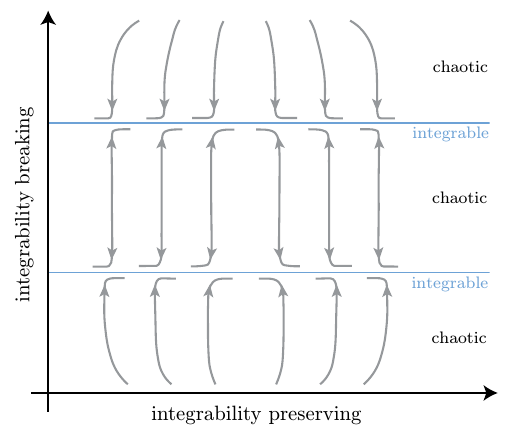}%
\caption[]{Schematic representation of the adiabatic flows in 2D parameter space. The two horizontal blue lines represent the lines of integrability while the regions outside them are chaotic. The horizontal (vertical) axis represents the integrability preserving (breaking) direction. By following the minimal $\chi$-directions, we can construct the shortest paths or flows (solid gray lines) that tend towards the integrable lines. The plot is based on the XXZ spin analyzed in this work.}
\label{fig:minimal-maximal}
\end{figure}

In this work, we analyze the properties of the QGT in two models, which have both integrable and chaotic regimes.  At each point in the (two-dimensional) coupling space, we diagonalize the QGT and find the two orthogonal directions that maximize and minimize the fidelity susceptibility. This procedure allows us to unambiguously define directions \textit{parallel} and \textit{orthogonal} to integrability when the model is integrable and smoothly extend these notions to chaotic regimes. We show that, for both directions corresponding to the minimal and the maximal values of the fidelity susceptibility, the corresponding observables conjugate to these couplings exhibit universal long-time dynamics. In particular, we find a good scaling collapse of the spectral functions of these observables in the chaotic but nonergodic regime, indicating universality of chaos close to integrability. Similar scaling collapse was found in a classical model~\cite{lim2024defining}. In this sense, integrable points share many similarities with critical points in continuous phase transitions: this emergent slow dynamics is analogous to critical slowing down.

The second result of the present work is the universality of the adiabatic flows near integrable regions. Specifically, we analyze the flows of the orthogonal directions diagonalizing the QGTs in the coupling space. Along these directions, the fidelity susceptibilities take their minimal and maximal values. Following the minimal directions, we can construct adiabatic flow lines similar to geodesics. Our key finding is schematically illustrated in Fig.~\ref{fig:minimal-maximal}. Namely, we observe that integrable regions are the attractors of these flows. In other words, if we follow the minimal $\chi$-directions, we will reach regions of integrability (if they exist) as the flows abruptly turn their directions to stay within the integrable regions. Physically, the minimal $\chi$-directions correspond to the fastest relaxation dynamics of the observables conjugate to these directions. In the model with an extended integrability breaking perturbation, these fast observables avoid long prethermalization. In the model with a boundary integrability breaking term, we find that prethermalization is longer along the direction parallel to the integrability. Hence, we can reformulate our result: following directions of fastest relaxation brings the system towards integrability. We provide both analytical and numerical evidence for the robustness of this statement beyond these specific models.

\section{\label{AGP}QGT and Fidelity Susceptibility}
In this section, we give a brief introduction to the concepts of the quantum geometric tensor and the fidelity susceptibility and discuss their regularization such that they are smooth, well-behaved functions of the couplings and system size. Much of the content here has been discussed in earlier papers (see Refs.~\cite{sugiura2021adiabatic,pandey2020adiabatic,kolodrubetz2017geometry,lim2024defining}), so we only mention details important for understanding the rest of this paper.

Suppose we have a Hamiltonian $\Hc(\vb*{\l})$ with coupling parameters $\vb*{\l}=\{\lambda_j\}$. The (eigenstate averaged) quantum geometric tensor~\footnote{In this paper, we are only interested in its real part also referred to as Fubini-study metric tensor} is defined as
\begin{equation}
    g_{ij}=\frac{1}{2\mathcal D} \sum_n \langle n| \mathcal A_i \mathcal A_j +\mathcal A_j \mathcal A_i |n\rangle_c,
\end{equation}
where $\Dc$ is the Hilbert space dimension~\footnote{for all subsequent calculations, we always consider only the central $50\%$ of the eigenstates}, $|n\rangle$ is the $\vb*{\l}$-dependent eigenstate of the Hamiltonian,  and the subindex ``c'' stands for the connected part or the covariance: $\langle n| \mathcal A_i \mathcal A_j |n\rangle_c\equiv 
\langle n| \mathcal A_i \mathcal A_j |n\rangle- \langle n| \mathcal A_i|n\rangle \langle n|\mathcal A_j |n\rangle$. Here, $\mathcal A_j$ is the adiabatic gauge potential (AGP) in the $j$-th direction, defined as the derivative operator acting on the eigenstates of the Hamiltonian (we set $\hbar=1$):
\begin{equation}\label{eqn:agp-propagate}
    i \partial_{\lambda_j} \ket{n(\vb*{\l})} = \Ac_j \ket{n(\vb*{\l})} \, ,
\end{equation}
The AGP and hence the QGT generally diverge in the thermodynamic limit and thus we consider their regularized version with the frequency cutoff $\mu$~\cite{pandey2020adiabatic}:
\begin{equation}\label{eqn:exact-agp}
    \mel{m}{\Ac_j}{n} = -i \frac{\o_{mn}}{\o_{mn}^2 + \mu^2} \mel{m}{\partial_j \Hc}{n} \, ,
\end{equation}
where $\o_{mn} \equiv \epsilon_m-\epsilon_n$ with $\epsilon_n$ and $\epsilon_m$ being the energy eigenvalues. In order to have a well-defined self-averaging behavior of the QGT with increasing system size, $\mu$ needs to be chosen to be parametrically larger than the typical level spacing. Then, the regularized QGT reads
\begin{equation}\label{eqn:agp-norm}
   g_{ij}=\frac{1}{\mathcal D} \sum_{n,m} \frac{\omega_{nm}^2}{(\omega_{nm}^2+\mu^2)^2} \langle m|\partial_i \mathcal H|n\rangle  \langle n|\partial_j \mathcal H|m\rangle.
\end{equation}
As in Ref.~\cite{pandey2020adiabatic}, we consider the range $1/\Dc_s \ll \mu \ll \Oc(1)$ \footnote{$1/\Dc_s$ is approximately the level spacing of the system.}, where $\Dc_s$ is the Hilbert space dimension of the largest symmetry sector of the model. At the same time, we choose $\mu$ to be smaller than other relevant energy scales in the problem such as the Thouless energy in the ergodic regime. Depending on the model, this scale can weakly (polynomially) depend on the system size $L$ while $\Dc_s$ typically depends exponentially on $L$ such that this range for $\mu$ is exponentially broad in $L$.

\begin{figure}[!t]
\centering
\includegraphics[width=8.6cm]{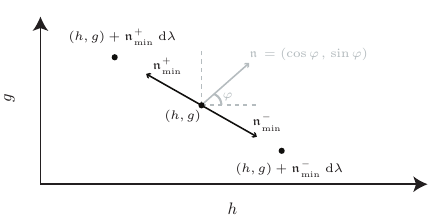}%
\caption[]{Graphical illustration of generic ($\mathfrak{n}$) and minimal ($\mathfrak n_{\rm min}$) directions. The gray portion denotes a generic direction $\mathfrak{n} = (\cos{\varphi}, \sin{\varphi})$. The adiabatic flows are obtained by following the minimal directions $\mathfrak n_{\rm min}^\pm$ starting at $\vb*{\lambda} = (h,g)$.}
\label{fig:flow-min-max}
\end{figure}

We focus on a two-dimensional control parameter space set by the couplings $h$ and $g$ (defined later for specific models) such that $\vb*{\lambda} = (h,g)$.  Let us observe that the AGP, as a derivative operator, transforms as a vector under rotations in the coupling space and hence the QGT transforms as a tensor. Suppose we have an infinitesimally small deformation such that $\vb*{\lambda}+{\mathfrak n}\, \dd \lambda  = (h+ \dd{\l} \cos{\varphi}, g+ \dd{\l} \sin{\varphi})$, where $\mathfrak n=(\cos{\varphi},\sin{\varphi})$ is the unit vector in the direction $\dd {\vb* \lambda}$. Then, the AGP along this direction is
\begin{equation}\label{eqn:chi-varphi}
    \Ac_{\mathfrak n}(\vb*{\l}) = \Ac_h \cos{\varphi} + \Ac_g \sin{\varphi} \, .
\end{equation}
Diagonalizing the QGT is therefore equivalent to finding the directions $\mathfrak n_{\rm min}$ and $\mathfrak n_{\rm max}$ defined by the angles $\varphi_{\rm min}(h,g)$ and $\varphi_{\max}(h,g)=\varphi_{\rm min}(h,g)+\pi/2$, respectively, along which the fidelity susceptibilities
\begin{multline}
\label{eqn:chi-min-max}
\chi_{\mathfrak n}\equiv g_{\mathfrak n\mathfrak n}=\frac{1}{\mathcal D}\sum_n \langle n|\mathcal A_{ n}^2|n\rangle_c\\
=\frac{1}{\mathcal D} \sum_{n,m} \frac{\omega_{nm}^2}{(\omega_{nm}^2+\mu^2)^2} |\langle m|( \mathfrak n\vdot{\vb * \nabla})  \Hc|n\rangle|^2 
\end{multline}
take their minimal and maximal values. Here ${\vb* \nabla}\equiv(\partial_h,\partial_g)$. Since the QGT has three independent entries, one can fully reconstruct it by computing $\chi_{\mathfrak n}$ along three different directions~\cite{kolodrubetz2017geometry}. Further, by \textit{continuously} following the minimal direction $\{ \mathfrak n_\text{min}(h,g) \}$,
we can reach a new eigenstate at $(h+\dd \lambda \cos{\varphi_{\rm min}} , g+ \dd \lambda \sin{\varphi_{\rm min}})$ and obtain the adiabatic flow diagram like in Fig.~\ref{fig:minimal-maximal}. Note that $\varphi_\text{min}$ is defined modulo $\pi$, corresponding to two anti-parallel directions $\mathfrak n_{\rm min}^+$ and $\mathfrak n_{\rm min}^-=-\mathfrak n_{\rm min}^+$ as shown in Fig.~\ref{fig:flow-min-max} (and so does $\varphi_\text{max}$). The precise choice of the direction $\mathfrak n_{\rm min}$ is not important as the geometric tensor is invariant under $\varphi \to \varphi+\pi$. We will return to this subtlety later when we demand continuity of the flows.

\section{The Models}

In this work we focus on analyzing two different models.
The first one is an XXZ spin chain of size $L-1$ with open boundary conditions coupled to a single spin at the boundary (coined as c-XXZ):
\begin{align} \label{eqn:XXZ}
\begin{split}
	\Hc_\text{c-XXZ} &= \frac{1}{2}\sum^{L-1}_{i=2} \left[\sigma_i^x \sigma_{i+1}^x + \sigma_i^y \sigma_{i+1}^y + \Delta \sigma_i^z \sigma_{i+1}^z \right] \\
    & + \frac{g}{2} \left(\sigma_1^x \sigma_{2}^x + \sigma_1^y \sigma_{2}^y + \Delta \sigma_1^z \sigma_{2}^z  \right) + h \sigma_1^z \, ,
\end{split}
\end{align}
where $\Delta$ is the anisotropy, $h$ is the local z-magnetic field strength on the spin at site $1$, and $g$ is the boundary exchange strength between the spin at site $1$ and the XXZ spin chain. For all subsequent calculations, we set $\Delta = 1.2$. Notably, when $g = 1$, we have an XXZ spin chain of length $L$ with a boundary magnetic field. Since an XXZ spin chain is integrable and a local field on the boundary does not break its integrability~\cite{alcaraz1987surface,gubin2012quantum}, this model is integrable along the line $g = 1$ for all $h$. Furthermore, the c-XXZ model is trivially integrable along the line $g = 0$ also for all $h$, and in the limits $h \to \infty$ and $g \to \infty$. At finite couplings in the regions outside of these two integrable lines, this model is expected to be chaotic as the $g$-exchange generically breaks integrability. 

For all subsequent calculations of the c-XXZ model (unless otherwise stated), we consider the $\expval{m} = 0, \pm 2$ ($\expval{m} = \pm 1$) magnetization sectors for even (odd) $L$, where $m \equiv \sum_i \sigma_i^z$. Following Ref.~\cite{pandey2020adiabatic}, we choose the cutoff $\mu =2L/\Dc_s$ for computing the QGT, where $\mathcal{D}_s = \binom{L}{\lfloor L/2 \rfloor}$ and $\lfloor L/2 \rfloor$ is the largest integer smaller than or equal to $L/2$. Physically, $\Dc_s$ corresponds to the dimension of the largest magnetization sector considered. This choice of $\mu$ allows us to study the limit $\mu\to 0$ with increasing $L$ while avoiding strong finite size effects due to finite level spacing. We refer to Appendix~\ref{appendix:finite-size} for further details on selecting the cutoff.

The second model we analyze is the Ising model with both transverse and longitudinal fields (LTFIM) with periodic boundary conditions
\begin{equation}
	\Hc_\text{LTFIM} = \sum^L_{i=1} \s_i^z \s_{i+1}^z + g \sum^L_{i=1} \s^x_i + h \sum^L_{i=1} \s^z_i \, ,
\end{equation}
where $g$ and $h$ are the transverse and longitudinal field strengths, respectively. This model is integrable along two lines: $h = 0$, as it maps to  free fermions via the Jordan-Wigner transformation, and $g = 0$, as it reduces to the classical Ising model. Finally, it is integrable at the point $h \to \infty$ and $g \to \infty$, as it becomes noninteracting. Outside of these regions, the model exhibits chaos~\cite{kim2014testing}. For all subsequent calculations (unless otherwise stated), we consider all $k \neq (0,\pi)$ quasi-momentum sectors. We use the cutoff $\mu =2.5 \cdot L/\Dc_s$, where $\mathcal{D}_s = 2^L/L$ and $\Dc_s$ is roughly the dimension of each $k \neq (0,\pi)$ quasi-momentum sector (see Appendix~\ref{appendix:finite-size} for more details).

\section{\label{spectral analysis} Spectral functions at and far from integrability}

Scaling of the fidelity susceptibilities with the frequency cutoff $\mu$ (and hence of the whole 
QGT) is determined by the spectral response~\cite{pandey2020adiabatic,lim2024defining}. To see this, observe that we can rewrite Eq.~\eqref{eqn:chi-min-max} as
\begin{equation}
\label{eq:g_through_Phi}
    \chi_{\mathfrak n}\equiv g_{\mathfrak n\mathfrak n} =\int^\infty_{-\infty} \dd{\omega} \frac{\omega^2}{(\omega^2+\mu^2)^2} \Phi_{\mathfrak{n}} (\omega) ,  
\end{equation}
where
\begin{equation}\label{eqn:spectral}
    \Phi_{\mathfrak n}(\o) = \frac{1}{\Dc} \sum_n \sum_{m \neq n} \abs{\mel{m}{ \partial_{\mathfrak n}   \Hc}{n}}^2 \delta(\omega - \omega_{mn})
\end{equation}
is the spectral function of the Hamiltonian deformation  in the direction $\mathfrak n\sim d{\vb* \lambda}$. Here, $\partial_{\mathfrak n}=(\mathfrak n \cdot \vb* \nabla)$ and  $\vb*\nabla= (\partial_h,\partial_g)$. In practice, we replace $\delta(x)$ with the Gaussian 
\begin{equation}\label{eqn:delta}
    \delta(x) \to \frac{1}{\Gamma \sqrt{2\pi}} e^{-x^2/(2\Gamma^2)}.
\end{equation}
We choose the broadening $\Gamma = \beta \,  \o_\text{H} > \o_\text{min}$, where $\o_\text{H}$ ($\o_\text{min}$) is the typical (minimum) level spacing of the central $50\%$ of eigenstates and we choose $\beta \sim 0.1 - 0.5$ depending on the extent of the finite-size effects.

From Eq.~\eqref{eq:g_through_Phi},
the behavior of $\chi_{\mathfrak n}$ is determined by the low-frequency asymptotes of the corresponding spectral functions (see also Refs.~\cite{leblond2021universality,bulchandani2022onset,pandey2020adiabatic,leblond2021universality,surace2023weak,orlov2023adiabatic,lim2024defining}). The focus of this work is understanding the structure of adiabatic flows and universality of the QGT and response close to integrability. It is instructive to first analyze the behaviors of the spectral functions in simpler limits where the system is either integrable or, conversely, is far from any integrable point. For this purpose, we numerically compute the spectral function $\Phi_{\mathfrak n}(\o)$ along two different directions for both models using exact diagonalization.

\begin{figure}[!htbp]
\centering
\includegraphics[width=8.6cm]{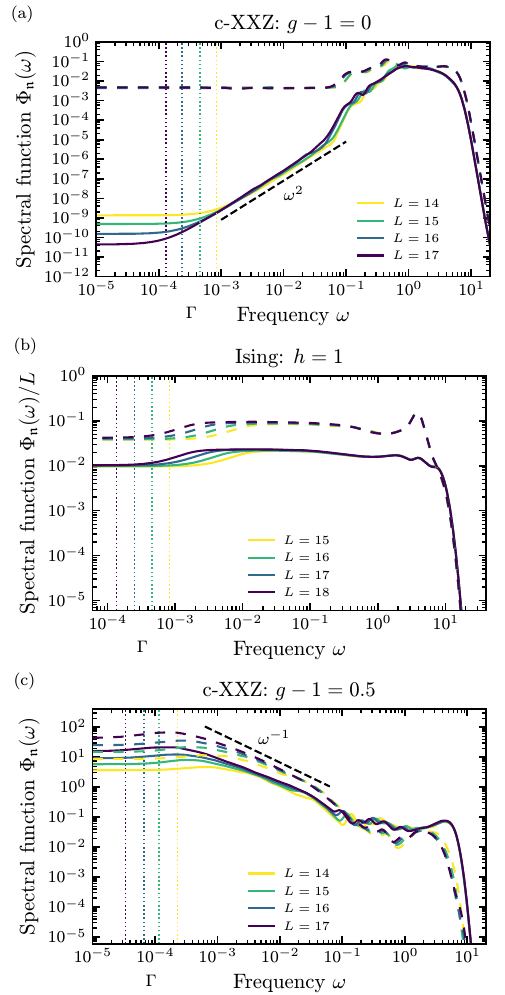}%
\caption[]{Spectral functions $\Phi_{\mathfrak n}(\o)$ at (a) integrable, (b) ergodic and (c) chaotic KAM regimes. (a) and (c) refer to the c-XXZ model at $h = 1.6$ with (a) $g-1=0$ and (c) $g-1=0.5$, respectively, for different system sizes $L = 14,15,16,17$. (b) refers to the nonintegrable Ising model with $(h,g) = (1,1.5)$ and system sizes $L = 15,16,17,18$. In (a), the solid [dashed] lines show $\Phi_{\mathfrak n}(\o)$ in the integrability preserving [breaking] directions: $\mathfrak n=(1,0)$ [$\mathfrak n=(0,1)$]. In (b) and (c), the solid (dashed) lines show $\Phi_{\mathfrak n}(\o)$ in the minimal (maximal) direction $\mathfrak n_{\rm min}\;(\mathfrak n_{\rm max})$. In (a) and (c), $\Phi_{\mathfrak n}(\o)$'s are averaged by taking $11$ and $3$ realizations of $h$ centered around $h = 1.6$ with maximum deviations of $5\%$ and $1\%$, respectively. In (b), $3$ realizations of $g$ centered around $g = 1.5$ with maximum deviation of $1\%$ are used. The vertical lines show the values of (a) $\Gamma = 0.5 \, \o_\text{H}$ and (b)--(c) $\Gamma = 0.1 \, \o_\text{H}$ used for different system sizes.}
\label{fig:xxz-ising-int-eth-kam}
\end{figure}

In Fig.~\ref{fig:xxz-ising-int-eth-kam}, we show examples of the spectral functions. Figure~\ref{fig:xxz-ising-int-eth-kam}(a) shows $\Phi_{\mathfrak n}(\omega)$ at the integrable point $\vb *\lambda=(1.6,1)$ of the c-XXZ model along the two natural directions: integrability preserving  $\mathfrak n=(1,0)$ and integrability breaking $\mathfrak n=(0,1)$, which correspond to changing the $h$ and $g$ coupling strengths, respectively. Figures~\ref{fig:xxz-ising-int-eth-kam}(b) and~\ref{fig:xxz-ising-int-eth-kam}(c) show the spectral functions of the (b) Ising model and (c) c-XXZ model far from integrable points--$\vb *\lambda=(1.5,1)$ and $\vb* \lambda=(1.6,1.5)$, respectively--along the minimal and maximal directions $\mathfrak n_{\rm min}$ and $\mathfrak n_{\rm max}$. Qualitative behaviors of these spectral functions agree with those observed earlier for other models. We briefly highlight some generic features and provide their interpretations.

For the integrable model [Fig.~\ref{fig:xxz-ising-int-eth-kam}(a)], we see that $\Phi_h(\o \to 0) \to 0$ while $\Phi_g(\o \to 0)$ saturates to a constant as it is generally expected~\cite{brenes2020low,leblond2021universality,pandey2020adiabatic}. The former indicates the existence of smooth adiabatic transformations of eigenstates and conservation laws as we change the integrability preserving coupling $h$. The saturation of $\Phi_g(\o)$ shows that the perturbation $\partial_g \mathcal H$ does not obey any selection rules, lifting any accidental degeneracies of the integrable model. We can observe (not shown) a similar qualitative behavior for the Ising model in the integrable line with a very sharp spectral gap along the $g$-line. 

Far from integrability, it is more natural to analyze $\Phi_{\mathfrak n}(\omega)$ along the minimal and maximal directions $\mathfrak n_{\rm min}$ and $\mathfrak n_{\rm max}$. In Fig.~\ref{fig:xxz-ising-int-eth-kam}(b), we show that the spectral functions along these directions are qualitatively similar to each other at low (and also high) frequencies. Far from integrability, this can be expected as generic observables should have similar long-time dynamics. The observed saturation of $\Phi_{\mathfrak n}(\omega)$ as $\omega\to 0$ agrees with the expected behaviors in RMT/ETH and can serve as an indication that the system is ergodic~\cite{pandey2020adiabatic,sels2021dynamical,leblond2021universality}.

Finally, in Fig.~\ref{fig:xxz-ising-int-eth-kam}(c), we present the spectral functions along the minimal and maximal directions for the c-XXZ model far from integrability, wherein low-frequency tails (at frequencies above the Heisenberg scale) develop along both directions. It is convenient to characterize these tails by the dynamical exponent $z$~\cite{sels2021dynamical}
\begin{equation}\label{eqn:spectral-z}
	\Phi_\mathfrak{n}(\o) = \frac{C}{\o^{{\rm max} (1-1/z,0)}}.
\end{equation}
It is thus clear that a close to $1/\omega$ tail observed numerically corresponds to the limit $z\gg 1$. Henceforth, we say that the spectral function has a low-frequency tail if it does not saturate at low frequencies except below the Heisenberg scale. A similar scaling was observed in disordered systems~\cite{sels2021dynamical} as well as classical chaotic models in the chaotic but not mixing regime~\cite{lim2024defining}. Since the existence of the latter in classical chaotic models is ensured by the Kolmogorov-Arnold-Moser (KAM) theorem~\cite{Chierchia:2010}, we will refer to this region as the KAM regime. This scaling of the spectral function saturates the upper bound of the scaling of the fidelity susceptibility with the cutoff $\mu$ and is characterized by a very slow power-law/logarithmic relaxation of the system to a non-thermal steady state. In Appendix~\ref{appendix:xxz-fidelity}, we analyze other indicators of ergodicity and come to a similar conclusion. While we are not able to extrapolate the spectral function to the thermodynamic limit, the results of the energy-level statistics are consistent with this KAM regime being transient like in disordered models~\cite{sels2017minimizing} and that, in the thermodynamic limit, this model could eventually become ergodic.

\begin{figure}[!htbp]
\centering
\includegraphics[width=8.6cm]{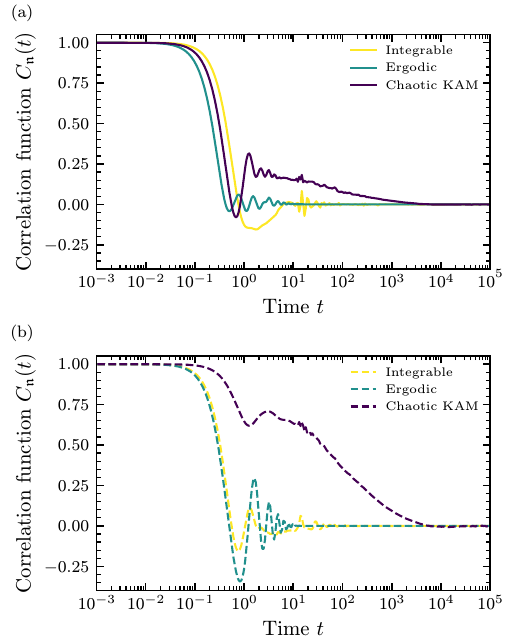}%
\caption[]{Symmetric correlation functions $C_{\mathfrak{n}}(t)$ at integrable, ergodic, and chaotic KAM regimes. We consider the same models and parameters for these regimes as done in Fig.~\ref{fig:xxz-ising-int-eth-kam} and only use the largest system size considered in each separate case. (a) and (b) depict $C_{\mathfrak{n}_\mathrm{min}}(t)$ and $C_{\mathfrak{n}_\mathrm{max}}(t)$ ($C_h(t)$ and $C_g(t)$ for the integrable case), respectively, with different regimes shown by separate colors.}
\label{fig:corr-int-eth-kam}
\end{figure}

We conclude this section by briefly translating these generic features of the spectral functions to the time domain by showing the corresponding behaviors of the symmetric correlation functions:
\begin{align}
\label{eq:C(t)}
C_\mathfrak{n}(t) &= \int^\infty_{-\infty} \dd{\omega} e^{-i\o t} \Phi_\mathfrak{n}(\omega) \, ,\\
&=  \frac{1}{2\Dc} \sum_{n}  \mel{n}{\poissonbracket{\partial_\mathfrak{n} \Hc(t)}{\partial_\mathfrak{n} \Hc(0)}}{n}_c \, ,
\end{align}
$\{ \ldots \}$ is the anticommutator, $\mel{n}{\partial_\mathfrak{n} \Hc(t) \partial_\mathfrak{n} \Hc(0)}{n}_c \equiv \mel{n}{\partial_\mathfrak{n} \Hc(t) \partial_\mathfrak{n} \Hc(0)}{n} - \mel{n}{\partial_\mathfrak{n} \Hc(t)}{n} \mel{n}{\partial_\mathfrak{n}\Hc(0)}{n}$, and
\begin{equation}
    \partial_\mathfrak{n} \Hc(t) \equiv e^{i\Hc t} \partial_\mathfrak{n} \Hc e^{-i \Hc t} \, .
\end{equation}
From these relations, it is clear that the low-frequency spectral behaviors correspond to the long-time relaxation dynamics.  We plot $C_{\mathfrak{n}_\mathrm{min}}(t)$ and $C_{\mathfrak{n}_\mathrm{max}}(t)$ ($C_h(t)$ and $C_g(t)$ for the integrable case) in Figs.~\ref{fig:corr-int-eth-kam}(a) and~\ref{fig:corr-int-eth-kam}(b), respectively, for the integrable, ergodic, and chaotic KAM regimes. We use the same parameters as in Fig.~\ref{fig:xxz-ising-int-eth-kam} corresponding only to the largest system size used. We observe a stark contrast between long-time relaxation dynamics in the chaotic KAM regime and the other two regimes. In the former, we see a slow relaxation of the correlation function that extends for several decades in time. This behavior reflects presence of the low-frequency $\sim 1/\omega$ tail in the spectral functions. On the other hand, we observe faster relaxation to the steady state in the integrable and ETH regimes.

\section{Universal dynamics and geometry close to integrability}\label{analysis}

\begin{figure}[!t]
\centering
\includegraphics[width=8.6cm]{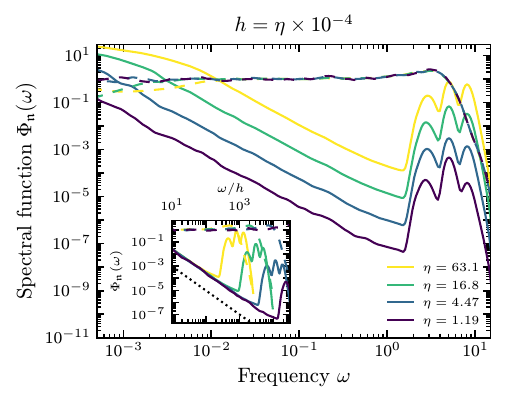}%
\caption[]{Spectral function $\Phi_{\mathfrak{n}}(\o)$ of the Ising model in the perturbative regime near integrability. We consider small integrability breaking perturbation strengths $h = \eta \times 10^{-4}$ for several $\eta$'s shown in different colors. The solid (dashed) lines refer to $\Phi_{\mathfrak{n}}(\o)$'s in the minimal (maximal) directions $\mathfrak{n}_\mathrm{min}$ ($\mathfrak{n}_\mathrm{max}$), which are nearly parallel to the $g$ ($h$) directions. The inset shows the perturbative scaling $\Phi_{\mathfrak{n}_\mathrm{min}} (\omega) \approx \Phi_g(\o) \sim h^2/\omega^2$, where the dotted line indicates this scaling collapse. Parameters: $L = 18$, $g = 1.5$ ($3$ different realizations of $g \in \{1.485, 1.5, 1.515\}$ are used for averaging), and $\Gamma = 0.5 \, \omega_\text{H}$.}
\label{fig:spec-ltfim-near-integrable}
\end{figure}

Having discussed some generic features of the spectral and correlation functions for these models at and far from the integrable regions, we will move to the main part of  our work. Namely, we will analyze universal properties of the dynamical response and of the QGT when the integrability breaking perturbation is small. It is intuitively clear that one can expect that both the response and QGT should be highly anisotropic, where they qualitatively behave differently between directions parallel and orthogonal to integrability.

Using Eq.~\eqref{eq:g_through_Phi}, we can estimate
\begin{equation}\label{eqn:agp-low-freq}
	\chi_{\mathfrak n}(\mu) \approx \begin{cases}
        \frac{\Phi_{\mathfrak n}(\mu)}{\mu} & \text{if } \mu > \Delta \, , \\
    \frac{\Phi_{\mathfrak n}(\Delta)}{\Delta} & \text{otherwise},
    \end{cases}
\end{equation}
where $\Delta$ is the spectral gap. From this result and Fig.~\ref{fig:xxz-ising-int-eth-kam}(a), we can conclude that, at the integrable lines, $\mathfrak n_{\rm min}$ coincides with the direction parallel to the integrability with $\chi_{\mathfrak n_{\rm min}}\sim {\rm const}(\mu)$ while $\mathfrak n_{\rm max}$ is the integrability breaking direction with $\chi_{\mathfrak n_{\rm max}}\sim 1/\mu.$ As we break integrability (consider the Ising model for concreteness) by adding a small longitudinal field $h$, we can anticipate that the spectral function along the $h$ direction does not change much as there are no selection rules for the matrix elements of $\partial_h \mathcal H$ even when $h=0$. Conversely, in the $g$ direction, it is easy to show using perturbation theory in $h$ that $\Phi_g(\omega)\sim h^2/\omega^2$, where we again use that, for the $h$-perturbation, there are no selection rules and all matrix elements are of the same order while the frequency $\omega$ sets the energy scale in the energy denominator (see Refs.~\cite{Garratt_2021,Garratt_2022} for a related discussion). To verify these perturbative results, we plot the spectral function $\Phi_{\mathfrak n}(\o)$ along the minimal and maximal directions for very small integrability breaking perturbation strengths in Fig.~\ref{fig:spec-ltfim-near-integrable} (see Fig.~\ref{fig:xxz-spec-g-1>0} in Appendix~\ref{appendix:spectral-analysis} for a similar plot in the c-XXZ model).

\begin{figure}[!t]
\centering
\includegraphics[width=8.6cm]{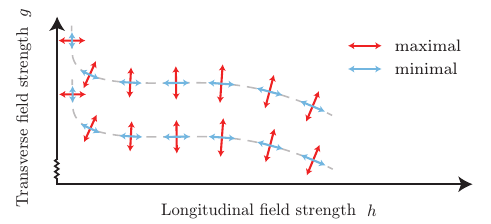}%
\caption[]{Schematic illustration of the structure of the QGT near an integrable line in the Ising model. The red and the blue arrows indicate directions $\mathfrak n_{\rm min}$ and $\mathfrak n_{\rm max}$, respectively, while the sizes of the arrows schematically (out of scale) indicate the magnitudes of the corresponding fidelity susceptibilities in these directions. The dashed lines illustrate the flows along the direction $\mathfrak n_{\rm min}$.}
\label{fig:flow-near-integrable}
\end{figure}

These perturbative considerations for the spectral function lead to the scaling predictions
\begin{equation}\label{eqn:ising-scaling}
    \chi_g \sim \frac{c h^2}{\mu^3} \quad \text{and} \quad \chi_h \sim \frac{c'}{\mu} \, ,
\end{equation}
with constants $c$ and $c'$. Comparing these two asymptotes, we come to an interesting conclusion: there exists a critical value of the integrability breaking perturbation $h_c\sim \mu$ such that $\chi_g=\chi_h$ and the minimal and the maximal directions switch depending on whether $h < h_c$ or $h > h_c$. Namely, in the regime where $h < h_c$, $\chi_g$ is smaller than $\chi_h$ and so $\mathfrak n_{\text{min}}$ ($\mathfrak n_\text{max}$) is aligned along the integrable (nonintegrable) direction, which is approximately parallel (perpendicular) to the line $h=0$. On the other hand, when $h > h_c$, the situation is reversed. In Fig.~\ref{fig:flow-near-integrable}, we give a pictorial representation of the minimal and maximal directions close to the integrable line $h = 0$. Note that as the cutoff $\mu$ gets smaller, corresponding to longer time scales, the switching crossover between the minimal and maximal directions becomes sharper and closer to the integrable line. Similar analysis applies to the c-XXZ model, such that at sufficiently large $L$ and small $\mu$ near the integrable line $g=1$, one expects
\begin{equation}\label{eqn:xxz-scaling}
    \chi_g \sim \frac{c}{\mu} \quad \text{and} \quad \chi_h \sim \frac{c'(g-1)^2}{\mu^3} \, ,
\end{equation}
with constants $c$ and $c'$ different from before.

\begin{figure*}[!htbp]
\centering
\includegraphics[width=2\columnwidth]{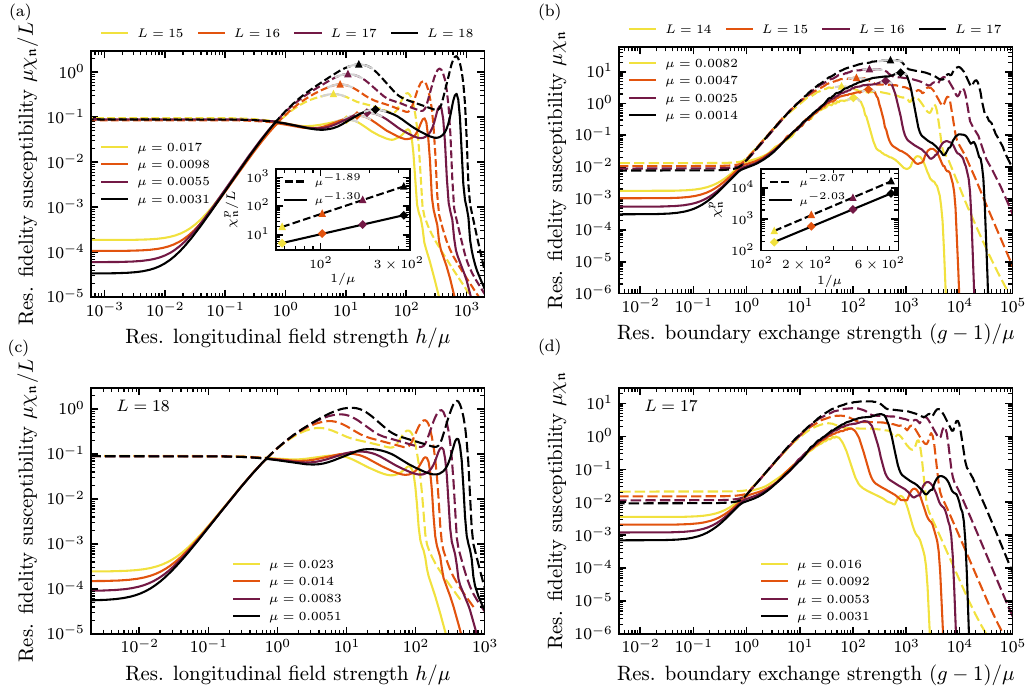}%
\caption[]{Scaling collapse of $\chi_{\mathfrak n}$ near integrability. In (a) and (c), we plot the rescaled fidelity susceptibility $\mu \chi/L$ against $h/\mu$ for the nonintegrable Ising model near the integrable line $h = 0$ with fixed $g = 1.5$. In (a), we vary $L$ and use $\mu = 2.5 \cdot L/\Dc_s$ with $\Dc_s = 2^L/L$ while, in (c), we fix $L = 18$ and vary $\mu$. In (b) and (d), we plot the rescaled fidelity susceptibility $\mu \chi$ against $(g-1)/\mu$ for the c-XXZ model near and above the integrable line $g=1$ at fixed $h = 1.6$. In (b), we vary $L$ and use $\mu = 2L/\mathcal{D}_s$, where $\mathcal{D}_s = \binom{L}{\lfloor L/2 \rfloor}$. In (d), we fix $L = 17$ and vary $\mu$. For (a)--(d), the solid (dashed) lines show fidelity susceptibility in the minimal (maximal) direction $\mathfrak n_{\rm min}\;(\mathfrak n_{\rm max})$. The insets in (a) and (b) show the peak values of $\chi_{\mathfrak n_{\rm min}}$ ($\chi_{\mathfrak n_{\rm max}}$) vs. $1/\mu$ as diamonds (triangles) with fits given by the solid (dashed) lines.}
\label{fig:agp-combined}
\end{figure*}

\begin{figure*}[!htbp]
\centering
\includegraphics[width=2\columnwidth]{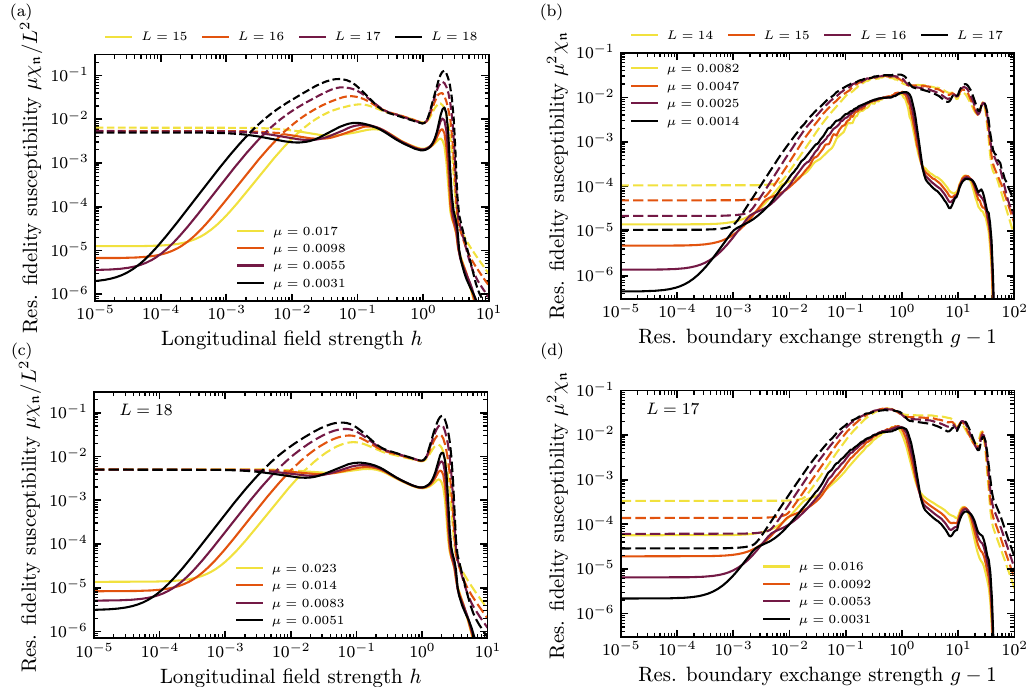}%
\caption[]{Scaling collapse of $\chi_{\mathfrak n}$ far from integrability. In (a) and (c), we plot $\mu \chi_{\mathfrak n} / L^2$ against $h$ for the nonintegrable Ising model with fixed $g = 1.5$. In (b) and (d), we plot $\mu^2 \chi_{\mathfrak n}$ against $g-1$ for the c-XXZ model at fixed $h = 1.6$. In (a) and (b), we vary $L$ and use $\mu = \alpha \cdot L/\Dc_s$ as defined in Fig.~\ref{fig:agp-combined}. In (c) and (d), we fix $L = 18$ and $L = 17$, respectively, and vary $\mu$. For (a)--(d), the solid (dashed) lines show $\chi_{\mathfrak{n}}$ in the minimal (maximal) direction $\mathfrak n_{\rm min}\;(\mathfrak n_{\rm max})$.}
\label{fig:chi-mu-combined}
\end{figure*}

The perturbative asymptotics for the fidelity susceptibilities in Eqs.~\eqref{eqn:ising-scaling} and~\eqref{eqn:xxz-scaling} suggest that, near integrable lines, they assume universal scaling forms:
\begin{align}
    \text{Ising:}& \quad  \chi_{\mathfrak n}=\frac{L}{\mu} f_{\mathfrak n}\left(\frac{h}{ \mu}\right)
    \label{eqn:ising-scaling-full} \, ,\\
    \text{c-XXZ:}& \quad \chi_{\mathfrak n}=\frac{1}{\mu} g_{\mathfrak n}\left(\frac{g-1}{\mu}\right) \, ,
    \label{eqn:xxz-scaling-full}
\end{align}
where $f_{\mathfrak n}(x)$ and $g_{\mathfrak n}(x)$ are direction-dependent scaling functions that recover perturbative results in the limit $x\to 0$. For the Ising model, we explicitly included the prefactor $L$ to account for the extensiveness of the deformations $\partial_h \mathcal H$ and $\partial_g \mathcal H$. To test this scaling hypothesis, we plot in Fig.~\ref{fig:agp-combined} the rescaled susceptibilities $\mu \chi_{\mathfrak n_{\rm min}}$ and $\mu \chi_{\mathfrak n_{\rm max}}$ (with an additional factor of $1/L$ for the Ising model) against the rescaled integrability breaking strength. Figures~\ref{fig:agp-combined}(a) and~\ref{fig:agp-combined}(c) correspond to the Ising model while Figs.~\ref{fig:agp-combined}(b) and~\ref{fig:agp-combined}(d) refer to the c-XXZ model. The top panels show the results for different system sizes, where the cutoff $\mu$ decreases with $L$ according to $\mu \propto L/\Dc_s(L)$, while the bottom panels correspond to a fixed $L$ with different lines describing different values of the cutoff $\mu$, which satisfy $1\gg \mu\gg 1/\mathcal D_s$. We see that for both models the fidelity susceptibilities develop maxima as functions of the integrability breaking parameters in agreement with Ref.~\cite{leblond2021universality}. On the left of these maxima, i.e., close to integrable regions, the fidelity susceptibilities in both directions exhibit the scaling collapse consistent with~Eqs.~\eqref{eqn:ising-scaling-full} and~\eqref{eqn:xxz-scaling-full}. A similar scaling collapse was also reported in a classical chaotic model at small integrability breaking~\cite{lim2024defining}, suggesting that the integrable regions in both quantum and classical models play a role similar to critical points for continuous phase transitions. It is interesting that, for both models, the scaling of the maximum of the susceptibility $\chi^p_{\mathfrak n_{\rm max}}$ is closely described by $\chi^p_{\mathfrak n_{\rm max}}\sim 1/\mu^2$ asymptote (see the insets in Fig.~\ref{fig:agp-combined}), in agreement with similar scaling observed in other models~\cite{sels2021dynamical,leblond2021universality}. This scaling saturates the upper bound of the divergence of $\chi$ with $\mu$ and can be termed as the region of maximal chaos or maximal complexity of the unitary transformation diagonalizing the Hamiltonian~\cite{sels2021dynamical,lim2024defining}. A similar $1/\mu^2$ scaling is observed for $\chi^p_{\mathfrak n_{\rm min}}$ for the c-XXZ model. The only exception is the scaling of $\chi^p_{\mathfrak n_{\rm min}}$ for the Ising model, which is approximately given by $1/\mu^{1.3}$. While we do not presently understand the origin of this anomalous scaling, we note that if we look into a generic direction $\mathfrak{n}$, then $\chi_{\mathfrak n}$ will be dominated by the more divergent direction and thus we expect to see $1/\mu^2$ scaling again for a generic direction $\mathfrak n$, as in Ref.~\cite{leblond2021universality}.

It is clear from Fig.~\ref{fig:agp-combined} that the scalings in Eqs.~\eqref{eqn:ising-scaling-full} and~\eqref{eqn:xxz-scaling-full} only work for sufficiently small integrability breaking perturbations, describing a crossover from the integrable to the KAM regime. At larger perturbations, we expect that, at least, the Ising model should become ergodic [see Fig.~\ref{fig:xxz-ising-int-eth-kam}(b)] such that the spectral function saturates in the regime $\omega\ll \omega_{\rm Th}$: $\Phi_{\mathfrak n}(\omega)\sim c$, where $c$ is some $h$-dependent constant and $\omega_{\rm Th}$ is the Thouless scale below which the energy spectrum can be described by RMT. Hence, we expect that $\chi_{\mathfrak n}(\mu)\sim c_{\mathfrak n}(h)/\mu$ in the regime where $\mu\lesssim \omega_{\rm Th}$. This motivates us to re-plot the rescaled $\chi_{\mathfrak n}$ shown in Fig.~\ref{fig:agp-combined} but, instead, against the unscaled integrability breaking perturbation $h$ as done in Figs.~\ref{fig:chi-mu-combined}(a) and~\ref{fig:chi-mu-combined}(c). As before, in Fig.~\ref{fig:chi-mu-combined}(a), we decrease $\mu$ with the system size while, in Fig.~\ref{fig:chi-mu-combined}(c), we fix the system size $L=18$ and lower $\mu$ while ensuring that it is always above the Heisenberg scale. In contrast to before, we have to account for an extra prefactor of $L$ for $\chi_\mathfrak{n}$ in the Ising model due to diffusion \footnote{For the Ising model at the ETH regime with varying system size, $\chi_\mathfrak{n} \sim c_\mathfrak{n}(h) L^2 /\mu$. One factor of $L$ trivially comes from the extensiveness of the observable. The remaining contribution comes the diffusive scaling $\Phi_\mathfrak{n}(\o) \sim 1/\sqrt{\omega}$ for $\o > \omega_\mathrm{Th}$ ($\Phi_\mathfrak{n}(\o) \sim 1/\sqrt{\omega_\mathrm{Th}}$ for $\omega < \omega_\mathrm{Th}$) with Thouless frequency $\omega_\mathrm{Th} \sim 1/L^2$, which adds an extra factor of $(1/L^2)^{-1/2} = L$ to $\chi_\mathfrak{n}$.}. At intermediate values of $h$, we see a very good data collapse indicating that the system is in the ergodic/ETH regime. The collapse region clearly tends to grow with increasing system size $L$ or, alternatively, time cutoff $1/\mu$. A similar collapse was observed in Ref.~\cite{leblond2021universality} for a different nonintegrable model. The situation is different for the c-XXZ model, where, according to Fig.~\ref{fig:xxz-ising-int-eth-kam}(c), even at strong integrability breaking, the spectral function develops a low-frequency tail, suggesting that the system is not ergodic, i.e., that $\omega_{\rm Th}=0$, at least for the available system sizes. The approximate $1/\omega$ behavior of this tail suggests that, for large values of integrability breaking perturbation $g-1$, the fidelity susceptibility should approximately scale as $\chi_{\mathfrak n}\sim c_{\mathfrak n}/\mu^2$, where $c$ is now a $g$-dependent constant. Indeed, this scaling agrees very well with the numerical results shown in Figs.~\ref{fig:chi-mu-combined}(b) and~\ref{fig:chi-mu-combined}(d) at intermediate values of $g-1$. As we mentioned already, this scaling saturates the upper bound of divergence of $\chi_{\mathfrak n}$ with the frequency cutoff $\mu$, suggesting that the model is always in the KAM regime at least for the accessible system sizes~\footnote{Note that $1/\mu^2$ scaling can be also explained if $\Phi_{\mathfrak{n}}(\omega)$ decays faster than $1/\omega$ but then there must exist a low frequency cutoff below which the spectral function must saturate~\cite{leblond2021universality}.}. We support this claim by observing the lack of collapse of $\mu\chi_{\mathfrak{n}}$ for any value of $g-1$, even for a different, extensive observable in Appendix~\ref{appendix:xxz-fidelity}. We note that the energy-level statistics results also shown in Appendix~\ref{appendix:xxz-fidelity} indicate that this model may eventually become ergodic in the thermodynamic limit. Extrapolation of the numerical data gives the lower bounds of the system sizes possible for entering into the ergodic/ETH regime, ranging from $L = 25$ (corresponding to $\mu \sim 7 \times 10^{-5}$) at large values of $g-1$ to $L \approx 180$ at a smaller value of $g-1=0.1$ (corresponding to $\mu \sim 10^{-50}$). The corresponding time cutoff $1/\mu$ sets the lower bound for the Thouless time required to observe thermalization in the model.

\begin{figure}[!t]
\centering
\includegraphics[width=8.6cm]{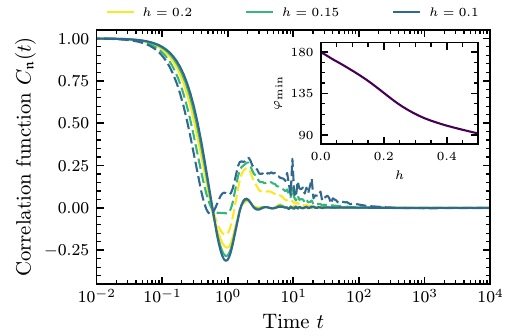}%
\caption[]{Relaxation dynamics of the observables in the nonintegrable Ising model for varying integrability breaking perturbation strengths. We plot the symmetric correlation function $C_\mathfrak{n}(t)$ in the minimal (solid lines) and maximal (dashed lines) directions for $L = 18$ at $g = 1.5$ with $h = 0.2, 0.15, 0.1$. Each line uses 3 different realizations of $g \in \{1.485,1.5,1.515\}$ for averaging. Inset shows $\varphi_\text{min}$ against $h$ near $h = 0$.
\label{fig:corr-function}}
\end{figure}

Let us contextualize the physical difference between the minimal and maximal directions in the regions sufficiently close to integrability but, simultaneously, in the non-perturbative regime. We focus on the nonintegrable Ising model, setting $g=1.5$ and the system size $L=18$, and vary the integrability breaking perturbation strength $h = 0.2,0.15,0.1$. These values of $h$ correspond to the vicinity of the maxima of $\chi_{\mathfrak{n}_\mathrm{min}}$ and $\chi_{\mathfrak{n}_\mathrm{max}}$ as shown in Fig.~\ref{fig:chi-mu-combined}(a). In Fig.~\ref{fig:corr-function}, we plot the corresponding symmetric correlation functions $C_\mathfrak{n}(t)$ [see Eq.~\eqref{eq:C(t)}]. We see that $C_{\mathfrak{n}_\mathrm{min}}(t)$ relaxes much faster than $C_{\mathfrak{n}_\mathrm{max}}(t)$, in agreement with Fig.~\ref{fig:corr-int-eth-kam}. This slow relaxation of $C_{\mathfrak{n}_\mathrm{max}}(t)$ physically corresponds to the long prethermalization of observables conjugate to the integrable direction. Perhaps surprisingly, we observe no evidence for prethermalization in the minimal direction $\mathfrak{n}_\mathrm{min}$ for these values of $h$. This is in contrast to the c-XXZ model, where both minimal and maximal directions show prethermalization as shown in Fig.~\ref{fig:corr-int-eth-kam}. As $h$ continues to approach zero, we observe that $C_{\mathfrak{n}_\mathrm{max}}(t) \approx C_g(t)$, where $\mathfrak{n}_\mathrm{max} \approx (0,1)$, decays at later times with increasingly noisy features marked by discrete many-body resonances due to finite level spacing~\cite{Bukov_2016,bulchandani2022onset} while $C_{\mathfrak{n}_\mathrm{min}}(t)$ does not change much. Further studies on the system size dependence of $C_\mathfrak{n}(t)$ at $h=0.1$ are shown in Appendix~\ref{appendix:spectral-analysis}. We thus see that the growing anisotropy between the minimal and maximal directions as one approaches the integrable line originates from drastically different relaxation dynamics between integrable and nonintegrable observables. We can thus conclude that the direction $\mathfrak{n}_\mathrm{min}$ ($\mathfrak{n}_\mathrm{max}$) physically corresponds to direction with the fastest (slowest) relaxation dynamics of the conjugate observables.

Let us point out that in Ref.~\cite{leblond2021universality}, contrary to this work, it was found that, even for a global integrability breaking perturbation, both integrability preserving and breaking directions showed similar scalings of the fidelity susceptibility with the system size for regions close to integrability. This difference between our and that work can be explained by the change of $\varphi_{\rm min}$ with respect to the integrability breaking coupling. We find that $C_{\mathfrak{n}_\mathrm{min}}(t)$ does not change much (for the Ising model) along this special direction $\mathfrak{n}_\mathrm{min}$. For any other direction $\mathfrak{n}\neq \mathfrak{n}_\mathrm{min}$, the observable $\partial_{\mathfrak n} \Hc$ always has nonzero overlap with $\partial_{\mathfrak n_{\rm max}} \Hc$ such that the long-time response is always dominated but the slow maximal direction. In particular, $\chi_{h}\approx \chi_{\mathfrak{n}_\mathrm{max}} \sin^2{\varphi_{\rm min}} +\chi_{\mathfrak{n}_\mathrm{min}} \cos^2{\varphi_{\rm min}} $. We find that $\varphi_\text{min}$ scales linearly with respect to $h$ near $h=0$ as shown in the inset of Fig.~\ref{fig:corr-function}.

\section{\label{numerical results}Flow Diagrams}

\begin{figure}[!htbp]
\centering
\includegraphics[width=8.6cm]{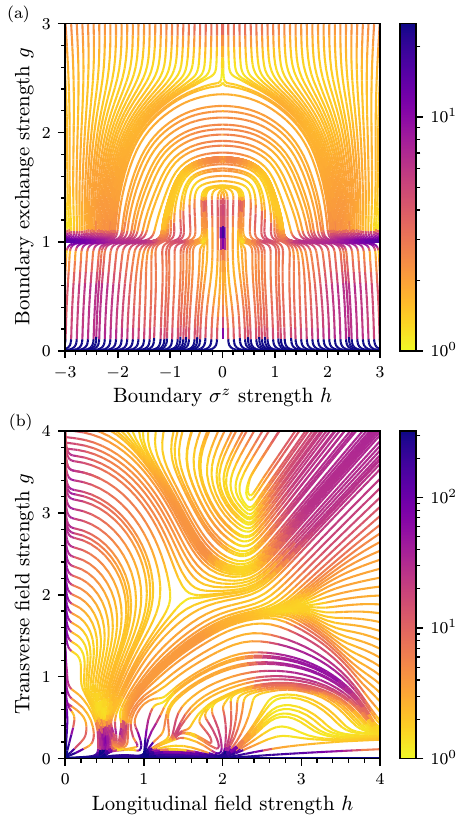}%
\caption[]{Flow diagrams. The paths that follow minimal directions $\mathfrak{n}_\mathrm{min}$ are presented for the (a) coupled XXZ model with $L = 18$ within the zero-magnetization sector and (b) nonintegrable Ising model with $L = 20$ within the $k = \pi/2$ quasi-momentum sector. Colors represent the anisotropy of $\chi_\mathfrak{n}$: the ratio of the $\chi_{\mathfrak{n}_\mathrm{max}}$ in maximal direction over $\chi_{\mathfrak{n}_\mathrm{min}}$ in the minimal direction. Along the line $g = 0$ for both models, the anisotropy strongly diverges and its value is over the limits of the color bars shown.}
\label{fig:flow-combined}
\end{figure}

Having understood the universal aspects of the geometric tensor near integrable regions, we present the full adiabatic flow diagrams. As explained earlier (see Fig.~\ref{fig:flow-min-max}) these diagrams are obtained by continuously following the minimal directions starting at various values of $\vb* \lambda$. As we explained in the text, these flows become universal and singular near the integrable lines. Away from these lines there are no singularities in the flow lines, which continuously drift with the system size or the frequency cutoff $\mu$. We thus do our best attempt to extrapolate the flow diagrams obtained using exact diagonalization to the thermodynamic limit.

\begin{figure}[!htbp]
\centering
\includegraphics[width=\columnwidth]{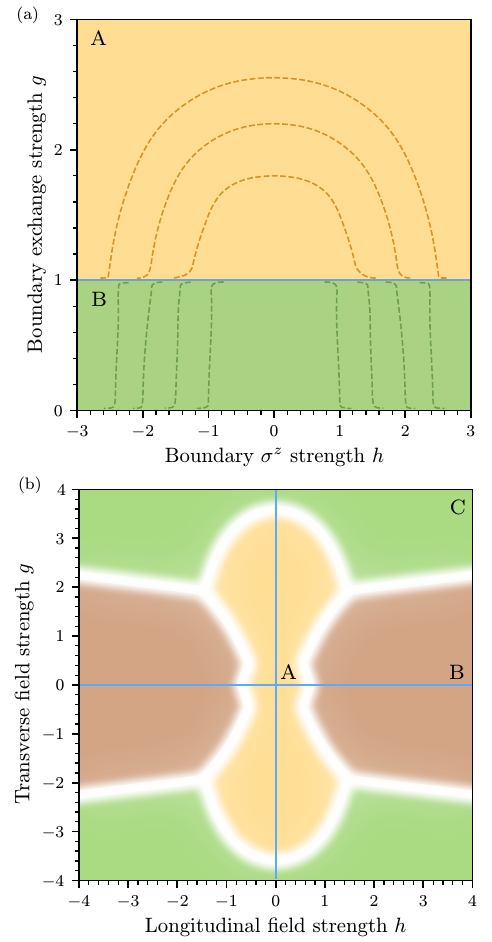}%
\caption[]{Phase diagrams. Qualitative pictures of the phases that connect regions of integrability (denoted by colored regions) of the parameter space are shown for the (a) coupled XXZ model and (b) nonintegrable Ising model, both in the thermodynamic limit. The phases are denoted by A, B, and C. The integrable lines are shown as blue lines while the dashed lines serve as guidelines to denote the connectivity of integrable regions within phases. The white lines in (b) signify the uncertainty of the phase separatrixes due to the limits of our numerical calculations.}
\label{fig:phase-combined}
\end{figure}

In Figs.~\ref{fig:flow-combined}(a) and~\ref{fig:flow-combined}(b), we show the flow diagrams for the c-XXZ model with $L = 18$ and the Ising model with $L = 20$, respectively. For the c-XXZ model, we consider the zero-magnetization sector, while we use the $k = \pi/2$ quasi-momentum sector for the Ising model. As done earlier for the c-XXZ model, we set $\mu = 2L/\Dc_s \approx 7.4\times10^{-4}$, where $\Dc_s = \binom{L}{L/2}$ is the dimension of the zero-magnetization sector, and for the Ising model $\mu = 2.5 \cdot L /\Dc_s  \approx 9.5 \times 10^{-4}$, where $\Dc_s = 2^L/L$ is the approximate dimension of the $k = \pi/2$ quasi-momentum sector. Remarkably, we can clearly identify integrable regions, without knowing them \textit{a priori}, as attractors of the flow lines, which become nearly orthogonal to the lines of integrability ($g = 0,\, g = 1$ for the c-XXZ model and $g = 0, \, h = 0$ for the Ising model) and then abruptly turn their directions after reaching integrability as illustratively shown in Fig.~\ref{fig:flow-near-integrable}. As explained earlier, these features become sharper with increasing system size and decreasing frequency cutoff $\mu$. As also observed in Figs.~\ref{fig:agp-combined} and~\ref{fig:chi-mu-combined}, we see that the anisotropy between the maximal and minimal directions increases (color darkens) near the integrable regions.

For the c-XXZ model, we can clearly identify two special points (vertices) on the integrable lines at $(h,g)=(0,0)$ and $(0,1)$. These are high degeneracy points, where, in addition to integrability, there are extra degeneracies due to the global $Z_2$ symmetry at $g=1$ and $Z_2\otimes SU(2)$ symmetry at $g=0$. Interestingly, we see that the $(0,1)$ vertex serves as an attractor of the adiabatic flows for $g<1$ but is repulsive for $g>1$. Similarly, $(0,0)$ is repulsive for $g>0$ and (though it is not shown) attractive for $g<0$. There are clearly also vertices at points $(|h|\to \infty,g=0)$ and $(|h|\to \infty,g=1)$, which can be also repulsive or attractive depending on which chaotic region they are in. For the Ising model, the flows near the whole integrable line $g=0$ become fragmented for $h\in (0,2)$ because this model has macrsocopic (exponential in the system size) degeneracies near rational values of $h$ in this interval, whose effects on the AGP were analyzed in Ref.~\cite{sugiura2021adiabatic}. Despite this fact, we can see that the line $g=0$ is still an attractor of the adiabatic flows. Moreover, as the system size increases, all singular behavior of the flows is pushed to lower values of $g$.

Now we do our best attempt to extrapolate the flow diagram to the thermodynamic limit by numerically examining trends of the flows for the coupled XXZ model with $L = 14, 16, 18$ and nonintegrable Ising model with $L = 16, 18, 20$, respectively (see Appendix~\ref{appendix:extrapolation} for more details on the extrapolation). We identify separatrixes that divide the coupling space into distinct sectors, where, in each sector, the flows terminate at different integrable lines. The results of this extrapolation are shown in Figs.~\ref{fig:phase-combined}(a) and~\ref{fig:phase-combined}(b). Note that there are no additional singularities near the separatrixes between different sectors except near integrable lines.

For the coupled XXZ model, there are two distinct sectors (denoted A and B) separated by the $g=1$ line that each connects different areas of integrability in the thermodynamic limit. Firstly, above the $g = 1$ line, we have region A of paths that connect points $(\pm h, 1)$ with curved arcs: that is, the shortest paths connect the integrable line $g = 1$ with itself. Below the $g = 1$ line, we have a distinct region B of almost vertical paths that connect the integrable lines $g =1$ and $g = 0$. Finally, below the line $g = 0$ line, there exists a region that connects the integrable line $g = 0$ with itself [not shown in Fig.~\ref{fig:phase-combined}(a)]. 

For the nonintegrable Ising model, we find three regions denoted A, B, and C in the thermodynamic limit. Region A is characterized by paths that connect the integrable lines $h = 0$ and $g = 0$, B by paths that connect the integrable line $g = 0$ with itself, and C by paths that connect integrable line $h = 0$ with either the integrable point $\sqrt{h^2 + g^2} \to \infty$ or line $g = 0$. Since extrapolating the precise form of the flows near the fragmented $g=0$ is difficult, we only extract the separatrixes between the regions, which are robust and do not change much with the system size.

\section{\label{conclusion}Conclusions and Outlook}

We found that integrable regions act as  attractors of the adiabatic flows. Specifically, the flows in the directions that minimize the quantum geometric tensor lead towards integrable regions. We showed that the underlying reason for this behavior is the parametrically faster relaxation of observables conjugate to these directions at small integrability breaking strength than for observables conjugate to directions parallel to integrability. 

We analyzed two one-dimensional models representing the coupled XXZ chain and the Ising model with boundary and bulk integrability breaking perturbations, respectively. For both models, we found that the geometric tensor exhibits universal scaling behavior near integrable lines, suggesting a close analogy between emerging chaos and continuous phase transitions. Further, we found strong numerical indications that the Ising model becomes ergodic and so satisfies ETH in the thermodynamic limit for any value of the integrability breaking perturbation. Conversely, we found that the coupled XXZ chain for any nonzero perturbation is in a chaotic KAM regime that does not obey ETH, at least for the available system sizes. Additionally, we numerically computed flow diagrams and identified distinct phases of flows that connect different or same areas of integrability and approximately extrapolated them to the thermodynamic limit.

There is an interesting corollary of our results, which suggests that systems, where external couplings are treated as macroscopic dynamical degrees of freedom, can generically self tune themselves close to integrable regions during autonomous time evolutions. Thus, ``integrability is attractive" not only in the mathematical sense but also as fixed points of time evolutions. This conclusion is in parallel to that of earlier works~\cite{Kofman_2004, Kolodrubetz_2015}, suggesting that, for systems close to the ground states, high symmetry or quantum critical points are natural attractors of dynamics. The physical reason for this dynamical attraction is that the divergent fidelity susceptibility comes with divergent dissipation and mass renormalization, leading to the freeze of time evolution along the directions with large $\chi_{\mathfrak n}$. Therefore, such dynamical systems would naturally evolve along the minimal directions.

Practically, our results pave the way to numerically finding nontrivial integrable or nearly integrable regimes either numerically by following the minimal directions $\mathfrak n_{\rm min}$ or experimentally by following directions with fastest long-time relaxation. They also suggest that it is plausible to develop a full scaling theory of emergence of chaos similar to the theory of continuous phase transitions. It is clear from our results that the full crossover between integrable and ergodic/ETH regimes can only be described by a two-parameter scaling theory.

\begin{acknowledgments}

This work was supported by the NSF Grant No. DMR-2103658 and the AFOSR Grant No. FA9550-21-1-0342. The numerical computations were performed using QuSpin~\cite{quspin1,quspin2}. The authors thank M. Rigol and L. Vidmar for useful comments on the manuscript and A. Dymarsky, M. Flynn, M. Rigol, and D. Sels for helpful discussions. The authors acknowledge that the computational work in this paper was performed on the Shared Computing Cluster administered by Boston University Research Computing Services.
\end{acknowledgments}

\appendix

\begin{figure*}[!htbp]
\centering
\includegraphics[width=2\columnwidth]{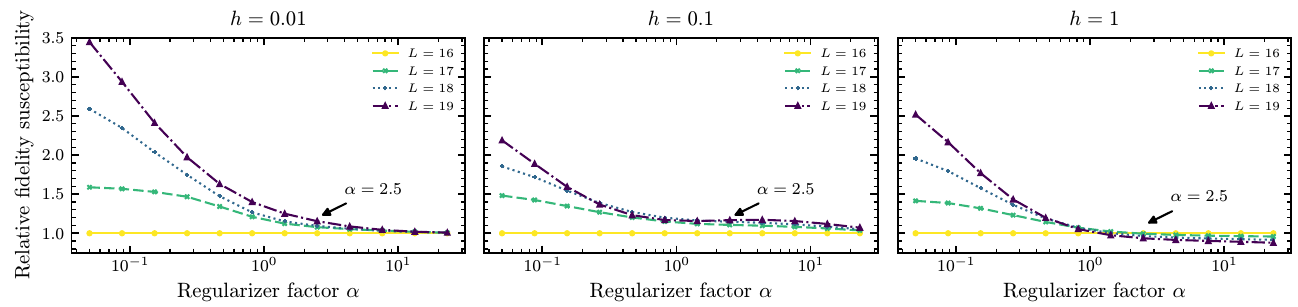}%
\caption[]{Cutoff dependence of the fidelity susceptibility. We plot the ratio $(\chi_h(L) / L) / (\chi_{h}(16) / 16)$ [$(\chi_h(L) / L^2) / (\chi_{h}(16) / 16^2)$ at ergodic regime $h = 1$] for the nonintegrable Ising model at $g = 1.5$ and $h = 0.01, 0.1, 1$ (from left to right). The cutoff $\mu$ is chosen to be the same for the different system sizes $L = 16,17,18,19$: $\mu = \alpha \cdot 16/\Dc_s(16)$, where $\Dc_s(L) = 2^L/L$. The arrow indicates the value $\alpha = 2.5$ used in the main text.}
\label{fig:agp-benchmark-ltfim}
\end{figure*}

\section{\label{appendix:finite-size} Cutoff optimization}

The exact computation of $\chi_{\mathfrak{n}}$ in Eq.~\eqref{eqn:chi-min-max} is not only dependent on parameters $\vb*{\lambda} = (h,g)$ but also the system size $L$ and the cutoff $\mu$. As discussed in the main text, for a given $L$, we use $\mu  = \alpha \cdot L/\Dc_s$, where $\Dc_s$ is the Hilbert space dimension and $\alpha$ is some constant with the goal of selecting $\alpha$ as small as possible while avoiding strong dependence on the system size from proximity to the Heisenberg scale. To motivate the choices of $\alpha$ used, we show the dependence of $\chi_{\mathfrak n}$ on $\alpha$. Specifically, we consider the nonintegrable Ising model at $g = 1.5$ for different values of $h = 0.01, 0.1, 1$. Then, we compute $\chi_h(L)/L$ (we divide by $L$ to remove extensiveness of the observable $\partial_h \Hc$) for system sizes $L = 16,17,18,19$ with a system size independent $\mu = \alpha \cdot 16/\Dc_s(16)$, where $\Dc_s(L) = 2^L/L$. In Fig.~\ref{fig:agp-benchmark-ltfim}, we plot the ratio of $\chi_h(L)/L$ to $\chi_h(16)/16$ [$(\chi_h(L)/L^2)/(\chi_h(16)/16^2)$ for $h = 1$] against $\alpha$. This is motivated from the scaling $\chi_h \sim L/\mu$ ($L^2/\mu$ for the ergodic regime $h=1$ when $\mu$ is below the Thouless frequency). The value of $\alpha=2.5$ used in the main text is highlighted with an arrow. Except for the smallest value of $h$ where finite size effects remain rather significant, we see that, with this choice of $\alpha$, the fidelity susceptibility $\chi_h(L)/L$ only depends on the cutoff $\mu$ and not on the system size. At $h = 1$, we see deviations when $\alpha$ is large due to closer proximity to the Thouless frequency $\omega_\mathrm{Th}$: when $\mu > \omega_\mathrm{Th}$, we expect the scaling $\chi_h \sim L/\mu^{3/2}$ instead.

\section{\label{appendix:xxz-fidelity}Nonergodicity of the c-XXZ model}

\begin{figure}[!t]
\centering
\includegraphics[width=8.6cm]{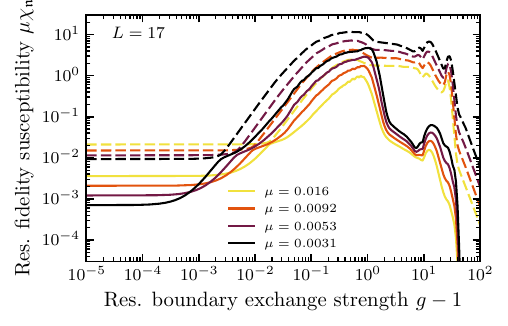}%
\caption[]{Fidelity susceptibility against boundary integrability breaking perturbation of the c-XXZ model. We plot the rescaled fidelity susceptibility $\mu \chi_{\mathfrak{n}}$ against $g-1$ near $g = 1$ at fixed $h = 1.6$ for system size $L = 17$. The solid (dashed) lines represent $\chi_\mathfrak{n}$ in the minimal (maximal) directions.}
\label{fig:xxz-chi-mu}
\end{figure}

\begin{figure}[!htbp]
\centering
\includegraphics[width=8.6cm]{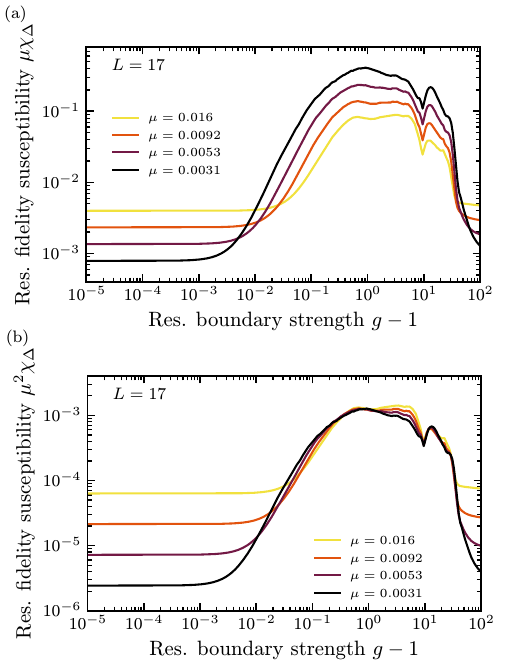}%
\caption[]{Fidelity susceptibility $\mathfrak \chi_{\Delta}$ corresponding to $\partial_\Delta \Hc=\sum_i \sigma_i^z \sigma_{i+1}^z$ against boundary integrability breaking perturbation of the c-XXZ model. (a) and (b) show $\mu \chi_\Delta$ and $\mu^2 \chi_\Delta$ against $g - 1$, respectively, at $h = 1.6$ for system size $L = 17$ with varying $\mu$.}
\label{fig:xxz-Delta-chi}
\end{figure}

As shown in Figs.~\ref{fig:chi-mu-combined}(b) and~\ref{fig:chi-mu-combined}(d), the fidelity susceptibility in the c-XXZ model scales as $1/\mu^2$, which is a signature of the maximally chaotic KAM regime. On the other hand, sufficiently far from integrability, $\chi_\mathfrak{n}$ scales as $1/\mu$ in the Ising model, which is an indicator of the ergodic ETH regime. In Fig.~\ref{fig:xxz-chi-mu}, we plot $\mu \chi_\mathfrak{n}$ against $g - 1$ at fixed $h = 1.6$ for the c-XXZ model with $L = 17$ and varying $\mu$. We do not observe any signs of persistent collapse of $\mu\chi_\mathfrak{n}$ as $\mu\to 0$ and hence the model cannot satisfy ETH for available system sizes (the Thouless energy must be less than the level spacing). We further support this claim by examining another observable conjugate to $\Delta$ with $\partial_\Delta \Hc=\sum_i{\sigma_i^z \sigma_{i+1}^z}$. Once again, we plot $\mu \chi_\Delta$ and $\mu^2 \chi_\Delta$ against $g-1$ in Figs.~\ref{fig:xxz-Delta-chi}(a) and~\ref{fig:xxz-Delta-chi}(b), respectively. Similar to $\chi_{\mathfrak{n}_\mathrm{min}}$'s and $\chi_{\mathfrak{n}_\mathrm{max}}$'s from the main text, $\chi_\Delta$ shows no indication of ergodicity and instead shows a chaotic KAM region that grows in size as $\mu \to 0$.

\begin{figure}[!htbp]
\centering
\includegraphics[width=8.6cm]{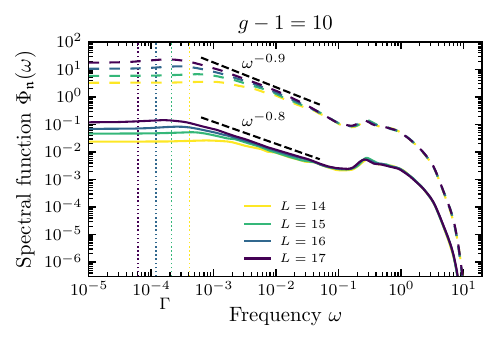}%
\caption[]{Spectral analysis of the c-XXZ model at large integrability breaking. We plot the spectral function $\Phi_\mathfrak{n}(\o)$ against frequency $\o$ at $g-1=10$ for system sizes $L = 14,15,16,17$. The solid (dashed) lines show $\Phi_\mathfrak{n}(\o)$ in the minimal (maximal) directions. Each line is averaged over $3$ realizations of $h \in \{1.584, 1.6, 1.616\}$. The vertical dotted lines show the values of $\Gamma 
= 0.1 \, \o_\text{H}$ used for different system size while the black dashed lines show the inverse frequency scaling of the spectral function: $\Phi_\mathfrak{n}(\o) \sim 1/\omega^\zeta$ with $\zeta \approx 1$.}
\label{fig:xxz-spec-chi-mu-squared}
\end{figure}

\begin{figure}[!htbp]
\centering
\includegraphics[width=8.6cm]{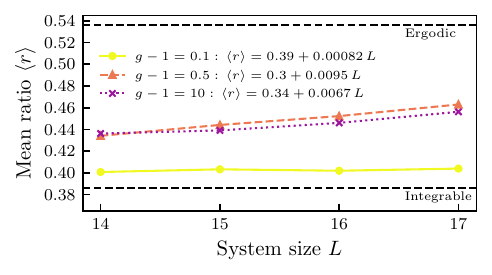}%
\caption[]{Level statistics of the coupled XXZ model. We plot the mean level spacing ratio $\expval{r}$ against system size $L$ at different integrability breaking coupling strengths $g-1 = 0.1, 0.5, 10$. $3$ different values of $h \in \{1.584, 1.6, 1.616\}$ are used for averaging. The horizontal dashed lines indicate the predictions for the Wigner-Dyson statistics (ergodic) $\expval{r} = 0.536$ and the Poisson statistics (integrable) $\expval{r} = 0.386$.}
\label{fig:xxz-mean-ratios}
\end{figure}

Next, we look closer to the spectral response for a fixed and very large integrability breaking perturbation strength $g - 1 = 10$ corresponding to the $\mu^2 \chi_\mathfrak{n}$ collapse regime in Figs.~\ref{fig:chi-mu-combined}(b) and~\ref{fig:chi-mu-combined}(d). In Fig.~\ref{fig:xxz-spec-chi-mu-squared}, we plot the spectral function $\Phi_{\mathfrak n}(\omega)$ for system sizes $L = 14,15,16,17$ at a fixed $h = 1.6$. We see that, for both Fig.~\ref{fig:xxz-ising-int-eth-kam}(c) (in the main text) and Fig.~\ref{fig:xxz-spec-chi-mu-squared}, $\Phi_\mathfrak{n}(\o) \sim 1/\o^\zeta$ at low frequencies with fixed $\zeta$ close to one, especially in the maximal directions. In turn, these low-frequency asymptotes of $\Phi_{\mathfrak n}(\o)$ lead to $\chi_{\mathfrak n}\sim 1/\mu^{1+\zeta}\approx 1/\mu^2$ scaling. The spectral functions do not show any saturation at low frequencies above the Heisenberg scale, which indicates that the Thouless energy is less than level spacing: for available system sizes, this model cannot satisfy ETH and hence is not ergodic.

We conclude this section by computing the mean level spacing ratio as a function of system size $L$, which is a more traditional measure for observing ETH. For each energy level spacing, $s_n = \epsilon_n - \epsilon_{n-1}$, the level spacing ratio is given as 
\begin{equation}
    r_n = \frac{\max(s_n, s_{n+1})}{\min(s_n, s_{n+1})} \, .
\end{equation}
Then, the value of the mean level spacing ratio $\expval{r}$ depends on the level statistics of the system: in the ETH/RMT regime, the system exhibits Wigner-Dyson statistics with $\expval{r} = 0.536$ while, in the integrable regime, the system exhibits Poisson statistics with $\expval{r} = 0.386$. A level statistics that gives $\expval{r}$ inbetween these two values is characteristic of the chaotic KAM regime. In Fig.~\ref{fig:xxz-mean-ratios}, we compute $\expval{r}$ for the c-XXZ model at specific points where we observe $\mu^2 \chi_\mathfrak{n}$ scaling collapse: $g-1 = 0.1, 0.5, 10$. Here, we vary system size $L$ and fix $h = 1.6$. We also show linear fits of $\expval{r}$ as functions of $L$. As expected, we see values of $\expval{r}$ that disagree with either ergodic or integrable behaviors. However, we observe that $\expval{r}$ tends to increase as $L$ increases. From the linear extrapolations, we conclude that $\expval{r}$ can reach the RMT value of $0.536$ approximately when $L = 25$ ($\mu \sim 7 \times 10^{-5}$) and $L = 30$ ($\mu \sim 2\times 10^{-6}$) when $g - 1 =0.5$ and $g - 1 = 10$, respectively. In contrast, we observe an extremely slow increase of $\expval{r}$ as a function of $L$ for $g - 1 =0.1$ and the linear extrapolation indicates that the system can become ergodic for $L \geq 180$ corresponding to $\mu \sim 10^{-50}$. These extrapolated values of cutoff $1/\mu$'s set the minimal values of the Thouless times $T_{\rm Th}=1/\mu$ that are required to observe thermalization for sufficiently large system sizes (where these times are shorter than the Heisenberg times). Let us point out that it is not certain that the linear increase of $\expval{r}$ with $L$ will not slow down as $L$ increases and that the system will thermalize at all in the thermodynamic limit. Even if it does thermalize, very large values of the Thouless time in the absence of small parameters in the model (i.e., when $g-1$ is large) look rather surprising.

\begin{figure}[!t]
\centering
\includegraphics[width=8.6cm]{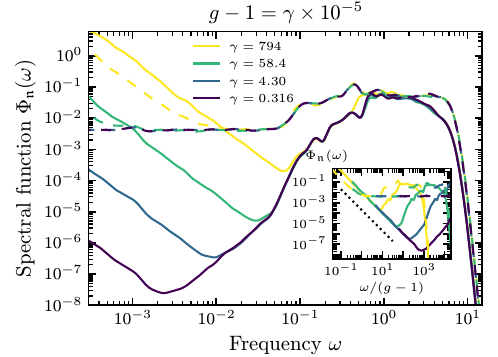}%
\caption[]{Spectral analysis of the c-XXZ model at small integrability breaking. We plot the spectral function $\Phi_\mathfrak{n}(\o)$ against frequency $\o$ for several values of $g - 1>0$ with system size $L = 17$. The solid (dashed) lines show $\Phi_\mathfrak{n}(\o)$ in the $h$ ($g$) direction. Each line is averaged over $3$ realizations of $h$ centered around $h = 1.6$ with maximum deviation of $1\%$. $\Gamma = 0.1 \, \o_\text{H}$ is used. Inset shows $\Phi_\mathfrak{n}(\o)$ against $\o/(g-1)$ while the dotted line indicates the $(\o/(g-1))^{-2}$ asymptote.}
\label{fig:xxz-spec-g-1>0}
\end{figure}

\begin{figure}[!b]
\centering
\includegraphics[width=8.6cm]{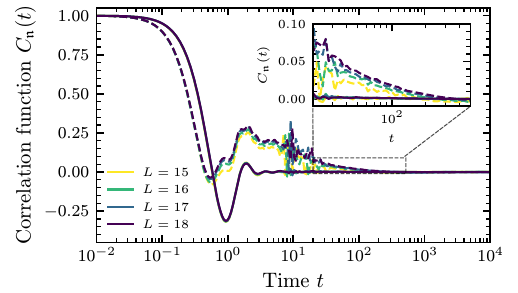}%
\caption[]{Symmetric correlation functions of the nonintegrable Ising model at $h = 0.1$ and $g = 1.5$ (same $g$'s for averaging as in Fig.~\ref{fig:corr-function}) for system sizes $L = 15,16,17,18$. We plot $C_\mathfrak{n}(t)$ against time $t$ while its inset shows a zoomed portion.}
\label{fig:corr-LTFIM-h-0dot1}
\end{figure}

\section{\label{appendix:spectral-analysis}Spectral analysis}

Here, we perform further analyses of the c-XXZ and nonintegrable Ising models to complement our discussion in Sec.~\ref{analysis}. 
Similar to Fig.~\ref{fig:spec-ltfim-near-integrable}, we plot the spectral function $\Phi_\mathfrak{n}(\o)$ close to the integrable point $(h = 1.6, g = 1)$ with system size $L = 17$ for the c-XXZ model in Fig.~\ref{fig:xxz-spec-g-1>0}. As shown, the low-frequency behaviors of $\Phi_\mathfrak{n}(\o)$ in both $g$ and $h$ directions are quite similar to each other as both directions show divergent spectral functions (for frequencies above the Heisenberg scale) at $\omega\to 0$. This is in stark contrast to those shown in the Ising model (refer to $\Phi_\mathfrak{n}(\o \to 0)$ in Fig.~\ref{fig:spec-ltfim-near-integrable}), where, along the integrability breaking direction, the spectral function remains flat.

\begin{figure}[!t]
\centering
\includegraphics[width=8.6cm]{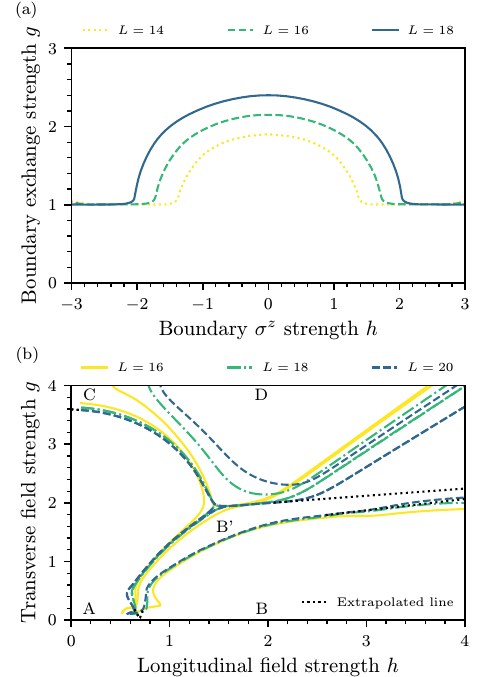}%
\caption[]{Separatrixes of flows. (a) shows the growth of phase region that connects the integrable line $g = 1$ with itself in the c-XXZ model for $L = 14,16,18$ and (b) shows the phase regions A, B, B', C, and D in the Ising model for $L = 16,18,20$ (see text for more details). The black dotted lines in (b) are the linearly extrapolated separatrixes.}
\label{fig:flow-extrapolation}
\end{figure}

We examine the system size dependence of $C_\mathfrak{n}(t)$ for the nonintegrable Ising model shown in Fig.~\ref{fig:corr-function}. We consider $h = 0.1$ and $g = 1.5$ for varying system sizes $L = 15,16,17,18$. We plot $C_\mathfrak{n}(t)$ against time $t$ in Fig.~\ref{fig:corr-LTFIM-h-0dot1}. This supports our findings that $C_{\mathfrak{n}_\mathrm{min}}(t)$ relaxes much faster than $C_{\mathfrak{n}_\mathrm{max}}(t)$ and that there is no sign of prethermalization in the direction $\mathfrak{n}_\mathrm{min}$. Further, we can observe that these conclusions hold true for the various system sizes considered. The inset of Fig.~\ref{fig:corr-LTFIM-h-0dot1} suggests that the long-lasting relaxation of $C_{\mathfrak{n}_\mathrm{max}}(t)$ extend to longer times as $L$ increases.

\begin{figure*}[!htbp]
\centering
\includegraphics[width=2\columnwidth]{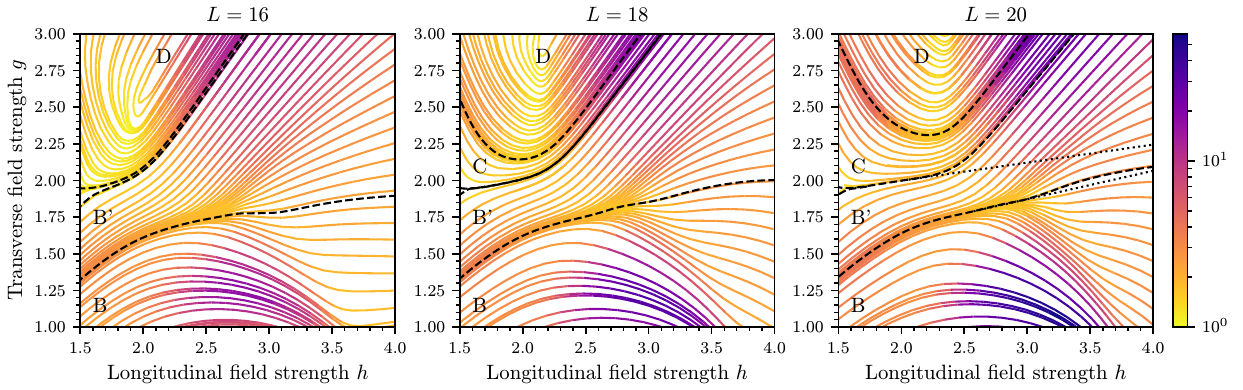}%
\caption[]{System size dependence of the flow diagrams in the nonintegrable Ising model. We plot the flow diagrams of the nonintegrable Ising model for system sizes $L = 16, 18, 20$ in the region $h \in [1.5,4.0]$ and $g \in [1.0, 3.0]$. The dashed lines represent the separatrixes between different regions B, B', C, and D while the dotted lines (shown for $L = 20$) represent the extrapolated lines of the separatrixes to the thermodynamic limit. The color bar represents the anisotropy of $\chi_\mathfrak{n}$.}
\label{fig:flow-16-18-20-zoomed}
\end{figure*}

\section{\label{appendix:extrapolation}Extrapolation of flow diagrams}

Here, we provide details on our numerical extrapolation of the infinite temperature phase diagrams in the thermodynamic limit $L \to \infty$.

For the c-XXZ model, to visualize the growth of region A as defined in Fig.~\ref{fig:phase-combined}(a), we examine the flow diagrams for system sizes $L = 14,16,18$. Here, we consider the zero-magnetization sectors with $\mu = 2L/\Dc_s$ and $\Dc_s = \binom{L}{L/2}$. We identify the outermost separatrixes of the semicircular flows above $g = 1$ and plot them for various $L$'s in Fig.~\ref{fig:flow-extrapolation}(a). As shown, the separatrix grows outwards without any signs of slowing down as $L$ increases and so we expect that the region above the line $g = 1$ to be dominated by region A in the thermodynamic limit.

For the Ising model, we identify five regions A, B, B', C, and D and then plot their separatrixes for system sizes $L = 16,18,20$ in Fig.~\ref{fig:flow-extrapolation}(b). Once again, we use all $k \neq (0,\pi)$ quasi-momentum sectors for $L = 16,18$ and the $k = \pi/2$ quasi-momentum sector for $L = 20$ with $\mu = 2.5 \cdot L/\Dc_s$ where $\Dc_s = 2^L/L$. Here, region D (B') denotes connectivity of the integrable point $I_\infty: \sqrt{h^2+g^2} \to \infty$ with itself (integrable line $g = 0$) [see Fig.~\ref{fig:flow-combined}(b) for visualization of regions]. The ``extrapolated lines'' denote the results of the linear extrapolations of separatrixes between regions A and C, B' and C, and B and B', respectively. For example, consider the separatrix between regions B' and C. As $L$ increases, there is a growing overlap of the separatrixes from different $L$'s towards larger $h$. We linearly extrapolate this overlap using the separatrix found at $L = 20$ in order to approximate its behavior in the thermodynamic limit. As shown, region A remains stable for $L \geq 18$ and region D tends to be pushed outward as $L$ increases. Further, the portion of region B' that connects to the integrable point $I_\infty$ gets pushed rightwards as $L$ increases as shown by the extrapolated line (also shown in the flow diagrams of Fig.~\ref{fig:flow-16-18-20-zoomed}). Then, in the thermodynamic limit, we expect region D to be nonexistent while regions B' to become part of region B. Therefore, we only expect three regions (A, B, and C) to survive in the thermodynamic limit as shown in Fig.~\ref{fig:phase-combined}(b). The expected survival of regions A, B, and C in the thermodynamic limit can be further examined by considering the flow diagrams for system sizes $L = 16,18,20$ in Fig.~\ref{fig:flow-16-18-20-zoomed}. We note that, while the accuracy of extrapolation to thermodynamic limit is only approximate, there are no singularities in the QGT away from the integrable regions. Therefore, the precise locations of the separatrixes are not universal and will depend on the details of the model. However, the general structure of the flow diagram is expected to be robust.

\bibliography{bibliography.bib}

\end{document}